\documentstyle[prl,aps,twocolumn]{revtex}
\begin{document}
\draft

\twocolumn[\hsize\textwidth\columnwidth\hsize\csname
@twocolumnfalse\endcsname
\title{A differential equation approach for examining the subtraction schemes}
\author{Ji-Feng Yang\thanks{email:jfyang@fudan.edu.cn}}
\address{School of Management, Fudan University, Shanghai 200433, P R China}
\date{June 16, 2001}
\maketitle

\begin{abstract}
We propose a natural differential equation with respect to
mass(es) to analyze the scheme dependence problem. It is shown
that the vertex functions subtracted at an arbitrary Euclidean
momentum (MOM) do not satisfy such differential equations, as
extra unphysical mass dependence is introduced which is shown to
lead to the violation of the canonical form of the Slavnov-Taylor
identities, a notorious fact with MOM schemes. By the way, the
traditional advantage of MOM schemes in decoupling issue is shown
to be lost in the context of Callan-Symanzik equations.
\end{abstract}

 ] \narrowtext

Conventionally, it is widely believed that different
renormalization schemes (that differ arbitrarily with each other)
are perfectly equivalent provided the perturbation is completed.
Once truncated at finite order, they yield the annoying scheme
dependence problem\cite{scheme}, i.e., different schemes yield
different predictions, and we have no better reasons to prefer one
scheme to the others. But recently, it has been realized that the
traditional on-shell definition of unstable particles masses and
width (say, e.g., masses and widths of $W^{\pm},Z^0$ and Higgs
particles in the electroweak model) is problematic and should be
replaced by the more physical and appropriate pole mass
schemes\cite{polem}, as the on-shell schemes would incur severe
gauge dependence and IR singularities or ambiguities. That means,
though we do not know the unique way of performing the physical
definition, the freedom in choosing renormalization conditions or
schemes deserves closer examination.

In this brief report, we suggest to examine the problem by
formulating the Feynman amplitudes as a series of differential
equations parametrized in terms of momenta (and masses for certain
theories)\footnote{Viewing the quantum field theories as just
effective sectors of the complete and well defined quantum theory
of everything (QTOE), we can derive finite loop integrals by
formulating them as solutions of appropriate differential
equations in terms of energy, momenta and masses, without
introducing any UV divergences, see Ref.\cite{Usa}.}, this is
feasible since the loop amplitudes must be functions of external
momenta and masses after all. Then, as we shall see shortly, not
all subtraction prescriptions or schemes satisfy such differential
equations. The differential equation analysis follows from the
well known fact that the differentiation with respect to physical
parameters like momenta and masses
lowers the divergence degree of a superficially divergent Feynman diagram%
\cite{Witten}. The solutions to these natural differential
equations must be well defined or finite functions of momenta (and
masses if any) and uniquely defined up to a certain set of
constants to be defined through 'boundary' conditions, which just
correspond to the definition of a renormalization/subtraction
scheme, the parameters of the solutions must have been finite
ones. It is worthwhile to point out that such analysis has already
applied in the usual renormalization programs through the critical
use of Ward identities, a set of partial differentiation equations
in terms of momenta. Our differential equation analysis is in fact
a mathematical generalization of the differential form Ward
identities. In this short report, we temporarily focus on the
differential equations in terms of masses for QFTs with massive
fields.

One might object such analysis with respect to masses as the
differentiation operation would shift or perturb the original
theory. Such an objection is simply unjustified. First, no matter
what kind of changes the differentiation brings into the theory,
it does bring us information between the two theories that
slightly differ in masses. Then as long as the difference is
essentially analytical, we can integrate the difference back to
find mass dependence of the original theory (up to ambiguities to
be fixed by boundary conditions). Second, a more direct and
mathematical way to see this is, if we believe that the theories
depend on the mass parameters in an almost analytical way, or that
the Feynman diagrams are regular functions of masses, then no
matter what the differentiation with respect to masses physically
brings about, we can perform differential analysis. Until now, we
have not learned of any example of QFT whose mass dependence is so
singular that it defies differentiation equation analysis. So, as
far as the mass dependence of Feynman amplitudes is concerned, it
does not matter whether or not a method shifted the original
theory or violated any symmetry of the original theory as long as
the perturbation or violation give rise to no completely singular
mass dependence. We do not know any special roles played by the
gauge symmetry and SUSY in preserving the analytical mass
dependence.

In fact such operation preserves most novel symmetries, like gauge
invariance, Lorentz invariance, and SUSY since canonical masses
are gauge invariant, Lorentz invariant and SUSY invariant if they
are masses of SUSY multiplet. For masses from symmetry breaking,
the differentiation equation analysis will become complicated due
to the complexity of the Higgs sector which will be examined in a
separate report. Here we focus on the invariant mass cases to
demonstrate that not all subtraction schemes are consistent with
the differential equation analysis. Note that we are not claiming
that these subtraction schemes can not consistently renormalize
QFTs with masses. We leave the implications of our result open to
the readers.

We will illustrate the analysis with a simple vertex function at
the lowest order in a QFT, say, the 1-loop photon vacuum
polarization tensor $%
\Pi ^{\mu \nu }(p,-p,m)$ ($\equiv -ie^2\int d^nk tr\{\gamma ^\mu \frac 1{%
p\!\!\!/+k\!\!\!/-m}\gamma ^\nu \frac 1{k\!\!\!/-m}\}$) in QED for
simplicity. But the main points of the analysis apply in the same way to
higher orders and to other theories like massive scalar theories and heavy
quark theories.

It is easy to see that this amplitude satisfies the following well defined
inhomogeneous differential equation in any gauge invariant regularization
scheme ($GIR$)
\FL
\begin{eqnarray}
&&\partial _m\Pi ^{\mu \nu }(p,-p,m) \nonumber \\
&=&-ie^2\int\left( d^4k\right) _{GIR}tr\{\gamma ^\mu (\frac
1{p\!\!\!/+k\!\!\!/-m})^2 \gamma ^\nu \frac 1{k\!\!\!/-m}
\nonumber \\
&+&\gamma ^\mu \frac 1{p\!\!\!/+k\!\!\!/-m}\gamma ^\nu (\frac 1{%
k\!\!\!/-m})^2\}
\end{eqnarray}
as the RHS of Eq.(1) is well defined in such schemes. Or it satisfies the
following well defined equation in any regularization schemes
\FL
\begin{eqnarray}
&&(\partial _m)^3\Pi ^{\mu \nu}(p,-p,m)=-ie^2\sum_{l=0}^3C_l^3
\int \left( d^4k\right) _{GIR} \nonumber \\ && tr\{\gamma ^\mu
(\frac 1{p\!\!\!/+k\!\!\!/-m})^{l+1}\gamma ^\nu (\frac
1{k\!\!\!/-m})^{4-l}\}
\end{eqnarray}
where $C_l^3$ is the combinatorial factor arising from the differentiation
operation.

It suffices to demonstrate our points with Eq.(1). Dropping the mass
independent transverse factor $(g^{\mu \nu }p^2-p^\mu p^\nu )$ we have the
following equation
\FL
\begin{equation}
\partial _m\Pi (p^2,m)=\frac{e^2}{\pi ^2}\int_0^1dx\frac{x(1-x)m}{%
m^2-x(1-x)p^2}.
\end{equation}
The solution to this equation is easy to find, it reads%
\FL
\begin{eqnarray}
\Pi \left( p^2,m;C\right)=&&\frac{e^2}{2\pi ^2}
\int_0^1dxx(1-x)\ln \frac{m^2-x(1-x)p^2}C\nonumber \\
  &+& F\left(
p^2\right)
\end{eqnarray}
with $C$ being the natural integration constant independent of to be fixed
by regularization and/or subtraction schemes and $F\left( p^2\right) $%
denoting an arbitrary mass independent function of momentum which can be
fixed to be zero by solving the differential equation in terms of momentum.
Then we have
\FL
\begin{eqnarray}
&&\Pi ^{\mu \nu }(p,-p,m;C)\nonumber \\
&=&\frac{e^2}{2\pi^2}(g^{\mu \nu }p^2-p^\mu p^\nu )\int_0^1dx\
(x-x^2)\ln \frac{m^2-(x-x^2)p^2}C\nonumber \\
\end{eqnarray}
with mass here being a finite or renormalized one. Note that the
constant $C$ here denotes any regularization scheme that is both
gauge invariant and mass independent. In other words, those
possible mass dependent regularization schemes are already
excluded, just like we exclude those regularization schemes that
violate gauge invariance and other physical symmetries in
realistic models.

In all mass independent schemes, the subtraction is so defined
that $C$ is fixed to be a mass independent constant which still
satisfies Eq.(3). However, in the MOM schemes, the subtraction is
defined in the following
way, 
\FL
\begin{eqnarray}
&&\Pi \left( p^2,m;C_{Mom}\right) \Vert _{p^2=-\mu ^2}=0 \nonumber
\\
&& \Longrightarrow C_{Mom}=m^2+x(1-x)\mu ^2
\end{eqnarray}
which ceases to be a solution of Eq.(3) but satisfies the following
equation
instead
\FL
\begin{eqnarray}
\partial _m\Pi \left( p^2,m;C_{Mom}\right) =&&\frac{e^2}{\pi ^2}\int_0^1dx%
\frac{x(1-x)m}{m^2-x(1-x)p^2}\nonumber \\ &+& \delta \left(
\mu^2,m\right) ,
\end{eqnarray}
with $\delta \left( \mu ^2,m\right) \equiv -\frac{e^2}{\pi ^2}\int_0^1dx%
\frac{x(1-x)m}{m^2+x(1-x)\mu ^2}$. It is clear in Eq.(7) that the
canonical relation of Eq.(3) is violated by the presence of {\it
an extra term that is dependent on the arbitrary subtraction point
$\mu $ }when massive particles are present. The extra term comes
from the nontrivial dependence of subtraction constant upon
particle mass in contrast to the mass independent schemes
(MIS)\cite{MIS} where $\partial _mC_{MIS}\equiv 0$, and therefore
becomes the source of problems in realistic applications.
Regarding the mass dependence, the on-shell scheme belongs to the
same class as MOM. It will lead to 'extra unphysical' mass
dependence for the vertex functions (including self-energies) for
theories with massive bosons, which is just the source of troubles
illustrated in \cite{polem}. The MOM like schemes are also known
to violate the canonical Ward or Slavnov-Taylor identities except
in the background gauge \cite{Reb}, which we think is related to
the fact pointed out above.

We would like to provide more arguments regarding the mass
dependence of renormalization schemes. Since the mass
differentiation would insert mass operators that is closely
related to the trace of the energy stress tensor which in turn
couples to gravity, such operation should not lead to any 'extra
unphysical' piece. It is easy to see that in Eq.(3) we can replace
$\partial_m$ by $m\partial_m$ without changing the problem.
Alternatively, the $m\partial_m$ operation is part of the overall
rescaling operation which also includes
$p^{\mu}\partial_{p^{\mu}}$. The latter operation should unveil
the physical dependence of vertex functions on spacetime, and
hence the mass dependence of vertex functions should also be
physical due to the overall rescaling entangle the mass dependence
with that of momenta. We can then understand why mass dependent
schemes often violate the canonical Ward identities, as the
differential form of Ward identities contain the operation
$\partial_{p^{\mu}}$, if we multiply this operator from the left
by $p^{\mu}$ then we see that it is part of the scaling equations.
Since the overall homogeneity of the vertex functions with respect
to all massive variables will entangle the momentum dependence
with the mass dependence, the extra term in Eq.(7) will lead to
violation of the canonical Ward identities. In this connection we
note that the pole mass for fermion is related to the effective
mass which is gauge invariant and IR finite\cite{Kronfeld,Tarrach}
by the equation $m_{eff}(p^2,{\bar
m})|_{p^2=M_{pol}^2}=M_{pol}$\cite{Tarrach}, i.e., the dependence
of pole mass on the MS running mass (Lagrangian mass) is {\sl
normal} since $m_{eff}$ can be defined in MS scheme though it is
scheme independent\cite{Coq}. The situation for bosons is similar,
see \cite{polem}.

Thus, contrary to the naive expectation, the differential equation
analysis does not allow for all kinds of subtraction schemes. That
means we should reexamine our conventional belief with due care.
If we accepted this approach, then the conventionally claimed
scheme invariance\cite{scheme} of certain objects within a subset
of schemes (i.e., within the mass independent schemes) can now be
viewed as a full 'invariance' for all schemes that solve the
differential equations in terms of all physical parameters.
Otherwise, the significance of the claims like the significance of
scheme invariance of the first two coefficients ($\beta _0,\beta
_1$) of beta function would be diminished by the mass dependent
schemes. In a sense we advanced a plausible rationale for using
mass independent schemes in our calculations. We also wish to
point out that our investigation in fact derives rationale from
the well known fact that not all regularization schemes would
satisfy the Ward identities. In our point of view, complicated
schemes (like MOM) not only bring about inconvenience, but also
might be unnatural choices. In other words, the freedom in
choosing the schemes is a restricted freedom if one works in the
differential equation approach suggested above.

Indeed there exists no literature up to date that provided convincing
constraints on mass dependence of schemes, but it is also hard to convince
people that a scheme that defines the one loop vacuum polarization tensor
like below would be a physically acceptable choice:
\FL
\begin{eqnarray}
&&\Pi ^{\mu \nu }(p,-p,m;\mu )\Vert _{Toy}\nonumber \\
&=&\frac{e^2}{2\pi ^2}(g^{\mu \nu }p^2-p^\mu p^\nu )\{
\int_0^1dxx(1-x)\ln \frac{m^2-x(1-x)p^2}{m^2}\nonumber \\ &+&
(\frac{m^2}{\mu^2 +m^2})^{1000}\},
\end{eqnarray}
that is to say, such a subtraction leaving a finite term like
$(\frac{m^2}{\mu^2 +m^2})^{1000}$ can not be an acceptable choice,
contrary to the belief that the subtraction can be arbitrary as
long as divergences are removed, at least the anomalous dimension
will be awkward-looking and very different from the standard
results. In fact we can consider the problem from the reverse
angle,i.e., suppose we found the true physical and hence scheme
independent parametrization, these physical parameters (together
with a RG and renormalization scheme invariant scale $\Lambda
_{phys}$) work coherently to give the unique mathematical formulas
that are often complicated functions. Then it is impossible to
transform these physical parameters into arbitrary things beyond
novel symmetries or invariances, i.e., it is impossible to
redefine the parameters in arbitrary schemes, since we must
confront our calculations with experiments and experiments will
only justify the schemes in which the dependence of the measured
quantities (defined as functionals of Feynman amplitudes) upon the
renormalized parameters are not quite different from the
dependence upon physical parameters, as the renormalized
parameters functionally take the same role as the physical ones,
an obvious fact in the formulation of QFT. Our investigation above
seems to add doubts to the conventional attitude to this issue.
Since it is just a first attempt, we will refrain from claim
anything, we only wish to draw attention to the above aspects.

The differential equations can be generalized to any order and any QFT with
massive fields: 
\FL
\begin{eqnarray}
&&(\partial_{m_i^s})^\omega {\Gamma
(p_1,\cdots,(m_i^s);[C])}\nonumber \\ &=&
\Gamma_{\frac{\phi_1^2}s,\cdots}
(p_1,{[0]}_\omega,\cdots,(m_i^s);{[C]}^{\prime}),\ \ \ \ s=1,2,\ \
\omega \geq 1,
\end{eqnarray}
where $(m_i^s)$ refers to the various masses in the theory with
$s=1$ for fermion and $s=2$ for boson, ${[C]}$ and ${[C]}^{\prime
}$ refer to the constants that will appear in the solutions and
$\omega $ being any times that will annihilate some of the
constants, which is the secret of the analysis in mass
differentiation.  Note that we have made use of the fact that the
differentiation with respect to mass(es) inserts the mass vertex
operators--the operator $(\phi _j^2/s)$ (for fermion it should be
understood as $\bar{\psi}\psi $).

Conventionally the MOM schemes are held to be advantageous over
the mass independent schemes in exhibiting good decoupling
behavior required by physics. In this connection, we wish to prove
that in the context of Callan-Symanzik equation\cite{CS} the
decoupling of heavy fields\cite{AC} is achieved in the same way in
both mass independent schemes and mass dependent schemes.
Especially, we do not need the so-called ''effective field
theories'' framework\cite{EFT} to help the mass independent
schemes, which is needed in the context of renormalization group
equation (RGE) since it does not account for the effects of {\it
all} the dimensional parameters. Such a proof is not seen yet in
literature. Again we will illustrate it with a simple model, QED
with a massive fermion in addition to $n_l$ massless fermions. In
a mass independent scheme the Callan-Symanzik equation reads,
\FL
\begin{eqnarray}
&&\{\lambda \partial_\lambda -\beta \alpha \partial _\alpha
+\gamma _\Gamma -D_\Gamma \}\Gamma ((\lambda p),m,\alpha ,\mu )
\nonumber \\ &=& -i\Gamma ^\Theta ((\lambda p),m,\alpha ,\mu )
\end{eqnarray}
where $\Theta \equiv {[1+\gamma_m]}m{\bar{\psi}}\psi $, $\beta $,
$\gamma _\Gamma $ and $\gamma _m$ are mass independent functions
of the renormalized coupling $\alpha $ and all quantities are
renormalized ones. At lowest order, $\beta =\frac{2\alpha }{3\pi
}{[n_l+1]}$.

When the mass goes to infinity, it is natural to expect that 
\FL
\begin{eqnarray}
&-&i\Gamma^\Theta ((p),m,\alpha ,\mu )|_{m\rightarrow \infty }
\nonumber \\ &=&(\Delta \beta \alpha \partial _\alpha -\Delta
\gamma _\Gamma )\Gamma ((p),\alpha ,\mu ),\\
&\rightarrow&\{\lambda
\partial_\lambda -\beta _{ls}\alpha \partial _\alpha +\gamma
_{\Gamma ;ls}-D_\Gamma \}\Gamma ((\lambda p),\alpha ,\mu
)|_{l.s.}=0.\nonumber \\
\end{eqnarray}
Here $\beta_{ls}\equiv \beta +\Delta \beta $, $\gamma _{\Gamma
;ls}\equiv \gamma _\Gamma +\Delta \gamma _\Gamma $ with the delta
contributions coming from the mass insertion part in the infinite
mass limit that will cancel the heavy fields' pieces in $\beta $
and $\gamma $. The generalization to other theories with boson
masses is an easy exercise. From Eq.(12) we see that the
decoupling of heavy particles is realized in a natural way in the
contexts of Callan-Symanzik equations.

To verify the above deduction it is enough to demonstrate Eq.(11)
at the lowest order which is closely related to the observation
that heavy particle limit provides a convenient
algorithm for calculating trace anomalies\cite{Trace}
\FL
\begin{eqnarray}
&&-i\langle m{\bar{\psi}}\psi J^\mu J^\nu \rangle |_{m\rightarrow
\infty }= \frac{2\alpha }{3\pi }(p^2g^{\mu \nu }-p^\mu
p^\nu)\nonumber \\&\Rightarrow& m(1+\gamma _m){\bar{\psi}}\psi
\Vert _{m\rightarrow \infty }=\frac 14\Delta \beta F^{\mu \nu
}F_{\mu \nu },\ \
\end{eqnarray}
with $\ J^\mu \equiv -ie{\bar{\psi}}\gamma ^\mu \psi $ and $\Delta \beta
\equiv -\frac{2\alpha }{3\pi }=\Delta \gamma _A$. When translated into
Callan-Symanzik equations Eq.(13) is just Eq.(11). The cancellation of the
heavy particle contributions is obvious since $\beta +\Delta \beta =\frac{%
2\alpha }{3\pi }{[n_l+1]}-\frac{2\alpha }{3\pi }=\frac{2\alpha
}{3\pi }n_l$.

While in the MOM schemes the Callan-Symanzik equation reads, 
\FL
\begin{eqnarray}
&\{&\lambda \partial _\lambda -\beta_{Mom}\alpha \partial _\alpha
+\gamma _{\Gamma;Mom}-D_\Gamma \}\Gamma_{Mom}((\lambda p),m,\alpha
,\mu )\nonumber \\ &=&-i\Gamma_{Mom}^\Theta ((\lambda p),m,\alpha
,\mu)
\end{eqnarray}
with the beta function,etc. being defined as $\ \beta_{Mom}\equiv
(\mu
\partial _\mu +m(1+\gamma_{m;Mom})\partial _m)\alpha ,\cdots ,$ in
contrast to the RGE definition: $\beta _{Mom}^{RG}\equiv \mu
\partial _\mu \alpha ,\cdots ,$ due to the mass dependence of the
renormalization constants\footnote{In the standard form of
Callan-Symanzik equation, the mass operators inserted vertex
functions appear separately and the definitions of $\beta,\gamma$
in the MOM like schemes must include $m(1+\gamma_{m;Mom})\partial
_m$ to account for the 'full running' of the relevant
renormalization constants. But in the alternative form of
Callan-Symanzik equation (CSE), i.e., $ \{\lambda \partial
_\lambda -\beta\alpha \partial _\alpha +m(1+\gamma_m)\partial
_m+\gamma _{\Gamma}-D_\Gamma \}\Gamma((\lambda p),m,\alpha ,\mu
)=0$, the definitions of the $\beta,\gamma$ etc. are the same as
in RGE. In this connection, we note that unfortunately many people
have confused CSE with RGE which are different things in
principle, CSE describes full scaling behavior while RGE only
states finite renormalization effects. In addition, in the
standard form of CSE, the trace of stress tensor is singled out,
which describes the physical coupling of matters to gravity and
therefore should be independent of renormalization schemes. Then
we can anticipate that the heavy particle limit of this term
should leave the same effects to light sectors in any
renormalization scheme, i.e., the same cancelling contributions to
the full $\beta, \gamma$ etc., just as what we will demonstrate
here.}.
The resulting $%
\beta_{Mom}$ also exhibits nondecoupling feature as in the mass
independent schemes. For example, at the lowest order, from the
definition
given above, a heavy particle's contribution to $\beta$ at lowest order is%
\FL
\begin{eqnarray}
\beta_{Mom}&=& (\mu \partial _\mu +m\partial _m)
\frac{e^2}{2\pi ^2}%
\int_0^1dxx(1-x)\ln C_{Mom}\nonumber \\ &=& \frac{2\alpha }{3\pi
}.
\end{eqnarray}
Then it is easy to find what is similar to Eq.(11)
\FL
\begin{eqnarray}
&-&i\Gamma_{Mom}^\Theta ((p),m,\alpha ,\mu )|_{m\rightarrow \infty
} \nonumber \\ &=&(\Delta \beta_{Mom}\alpha \partial _\alpha
-\Delta \gamma_{\Gamma;Mom})\Gamma_{Mom}((p),\alpha ,\mu)|_{ls},
\end{eqnarray}
with $\Delta \beta_{Mom}=-\frac{2\alpha }{3\pi }$ as Eq.(13) is
also true in the MOM schemes at the lowest order. It is known that
the first loop order beta of RGE in mass independent schemes
differs from that in the MOM schemes\cite{Coq}. While in the
context of Callan-Symanzik equation the first loop order beta
function is the same in all schemes at one loop order (C.f.
Eq.(15)), which in turn implies that the same decoupling mechanism
works in MOM schemes in the context of Callan-Symanzik equation.
Thus the advantage of MOM schemes over the mass independent
schemes in the decoupling issue is lost in the context of
Callan-Symanzik equation and the decoupling is realized in the
same way specified by Eq.(11).

We should note that, many MOM scheme calculations of couplings
\cite{Jeg} are in fact calculations of the effective couplings
with the momentum transfer defined at an Euclidean point. The CWZ
\cite{CWZ} scheme for calculations in the presence of heavy quarks
is in fact a $\overline{MS}$ scheme with the running scale fixed
to be the particle mass, since in this
scheme the mass also serve as a running scale for massive loops and the $%
\beta $ function, etc. are all mass independent. We should also
note that our analysis only shows that the mass dependent schemes
like MOM and on-shell schemes {\sl might} cause problems for {\sl
massive and some related sectors} (like self-energy vertices,etc.)
in a QFT. For massless sectors or massless QFTs, the problem does
not materialize.

If one adopts our proposal for renormalization mentioned
above\cite{Usa}, then the decoupling of heavy particles can be
achieved in an automatic way as our proposal stands on the
philosophical ground of the effective field theories\cite{EFT},
please see Ref. \cite{CSE}.

In summary, we suggested a differential equation analysis of the
radiative corrections in terms of momenta and masses and
demonstrated that only certain subtraction schemes satisfy such
equations. Some conventional schemes like MOM failed to solve
these equations. By the way, we gave a demonstration that in the
context of the Callan-Symanzik equations all the schemes
facilitate the decoupling of heavy particles in the same way,
weakening the argument that the MOM like schemes is superior to
mass independent schemes in decoupling issue.

\end{document}